\documentclass[%
reprint,
 amsmath,amssymb,
 aps,twocolumn, 
]{revtex4-1}

\usepackage{float}
\usepackage{graphicx}% Include figure files
\usepackage{dcolumn}% Align table columns on decimal point
\usepackage{bm}% bold math
\usepackage[flushleft]{threeparttable}
\usepackage{url}
\usepackage{amsmath}
\usepackage{amssymb}
\usepackage{multirow}
\usepackage{array}
\usepackage{ulem}
\newcolumntype{P}[1]{>{\centering\arraybackslash}p{#1}}
\usepackage{xcolor}
\usepackage{framed}
\usepackage{lipsum}
\colorlet{shadecolor}{yellow!20}
\makeatletter
\newcommand*{\rom}[1]{\expandafter\@slowromancap\romannumeral #1@}

\begin{document}

\preprint{APS/123-QED}

\title{``Double-path'' ferroelectrics and the sign of the piezoelectric response}% Force line breaks with \\

\author{Yubo Qi$^1$, Sebastian E. Reyes-Lillo$^2$, and Karin M. Rabe$^1$}

\affiliation{%
$^1$Department of Physics $\&$ Astronomy, Rutgers University, 
Piscataway, New Jersey 08854, United States\\
$^2$Departamento de Ciencias F\'{i}sicas, Universidad Andres Bello,
Santiago 837-0136, Chile
}%

\pacs{Valid PACS appear here}

\begin{abstract}

In this work, we propose a class of ferroelectrics (which we denote ``double-path'' ferroelectrics), characterized by two competing polarization switching paths for which the change in polarization is different and in fact of opposite sign.
Depending on which path is favorable under given conditions, this leads to different identification of up- and down-polarized states. 
Since the sign of piezoelectric response depends on the assignment of up- or down-polarized state for a specific structure,
this means that the material can exhibit different signs of the piezoelectric response under different conditions.
We focus on HfO$_2$ as a key example.
Our first-principles calculations show that there are two competing paths in HfO$_2$, resulting from different displacements of the atoms from the initial to the final structures, and the change in polarization along these two paths is of opposite sign. 
These results provide a natural explanation for the recently observed discrepancy in the signs of piezoelectric responses in HfO$_2$ between theoretical first-principles calculations and experimental observation. Further, this allows predictions of how to favor one path over another by changes in conditions and compositional tuning.
This family of materials also includes other candidates, such as CuInP$_2$S$_6$ and theoretically proposed LaVO$_3$-SrVO$_3$ superlattice.
We finally note that double-path ferroelectrics possess novel electromechanical properties since the signs of their piezoelectric responses can be switched.

\end{abstract}

\maketitle

Ferroelectrics have promising applications in electronic devices due to their spontaneous polarization and switchability by applied electric fields~\cite{Haertling99p797,Setter06p051606}.
Typically, a ferroelectric material has two symmetry-related states, called variants, with polarization in opposite directions, designated ``up'' and ``down'', with switching from the up state to the down state by application of a field in the down direction, and from down to up by a field in the up direction. The ferroelectric polarization is obtained as half of the magnitude of change in polarization on switching. 

In conventional ferroelectrics such as BaTiO$_3$ or PbTiO$_3$, the polar states are generated by freezing in of an unstable polar lattice mode of a closely related high-symmetry state, such as the cubic perovskite structure~\cite{Ghosez99p836,Sai02p104108,Dieguez05p144101,Qi16p134308,Qi15p245431}.
In that case, the identification of a particular variant as up or down polarized is straightforward. 
The displacement of each ion in the up state to its final position in the down state is determined by reversing the displacements of the unstable polar lattice mode distortion, with the overall net displacement along the direction of the electric forces on the ions.
This matching up corresponds to a switching path with displacements that are in some sense minimal, and from this path, the change in polarization can be directly computed~\cite{Bonini20p045141}.

\begin{figure}[hpbt]
\includegraphics[width=8.5cm]{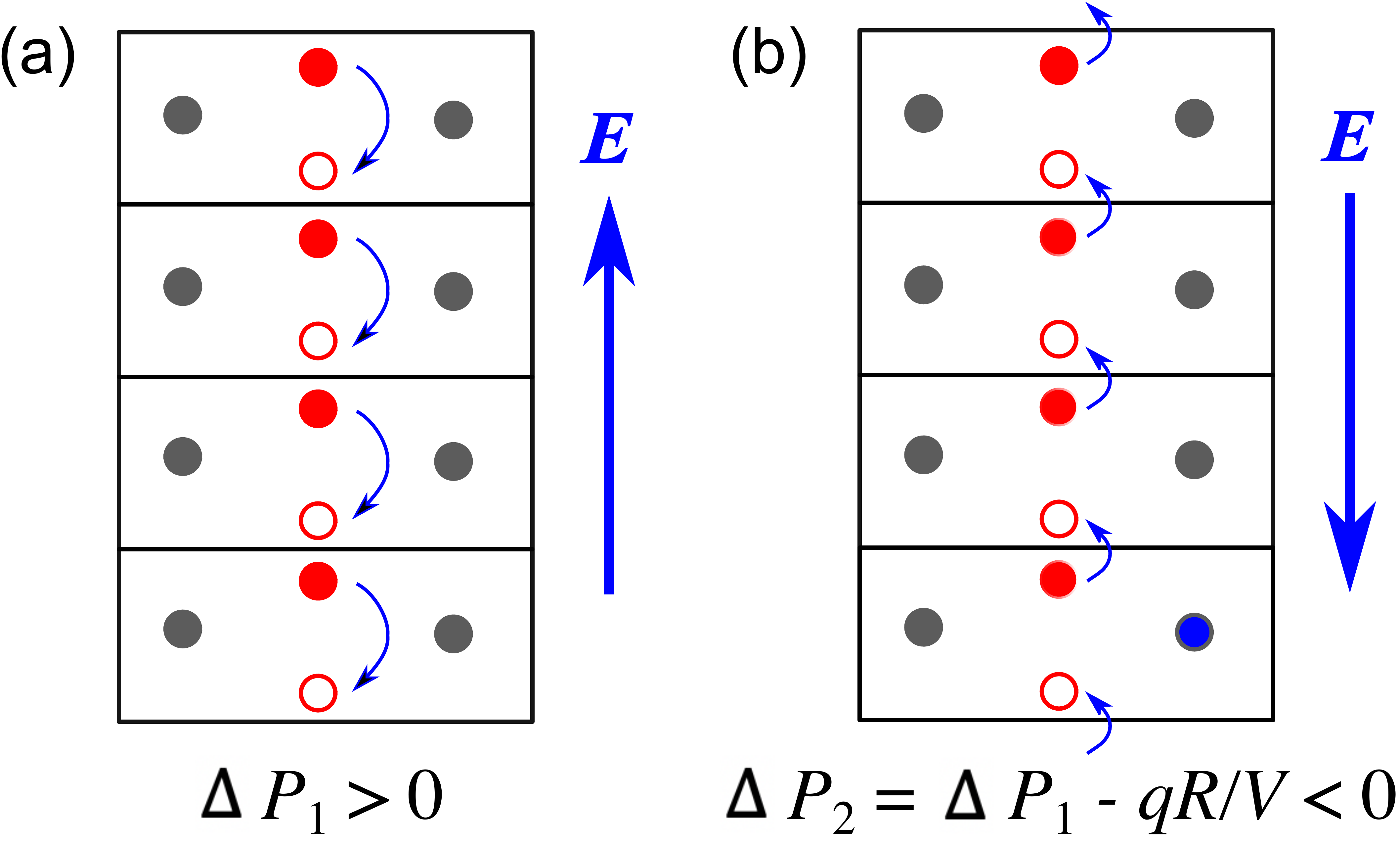}
\caption{Schematic illustrations of path dependent switching polarization. Filled red and grey circles represent negatively and positively charged ions in the crystal. The arrows indicate the motion of the red ions in the applied electric field shown, and open red circles represent the final positions of each red ion after switching. }
\label{f1}
\end{figure}

In some ferroelectrics, the matching up of ions in the up state with final positions in the down state can be made in two different ways with comparable ionic motion, as illustrated in Fig.~\ref{f1}.
The initial structures in the two subfigures (represented by the grey circles and red solid circles) are the same and the final structures in the two subfigures (represented by the grey circles and red hollow circles) are also the same.
However, due to the different displacements of the red cations, the switching polarizations in Fig. 1 (a) and (b) differ by a integer multiple of the quantum of polarization $q{\bm{R}}/V$, where $q$ is the ionic change, $\bm{R}$ is the lattice vector, and $V$ is the volume~\cite{Bonini20p045141,King93p1651}, and in fact are opposite in sign. The switching in (a) is thus driven by an upward-pointing electric field, as shown, leading to the identification of the initial structure as down polarized. Conversely, the switching in (b) is driven by an downward-pointing electric field, as shown, leading to the identification of the same initial structure as up polarized.

In this work, we argue that ferroelectric HfO$_2$ is a double-path ferroelectric. Combining previous results~\cite{Clima14p092906,Clima20p1,Yang20p064012,Huan14p064111,Maeda17p1,Fan19p21743,Choe21p8,Chen20p085304,Ding20p556,Lee20p1343} with additional first-principles calculations, we will show that there are competing paths connecting symmetry related variants that match the ions in one state to final positions in the other state in different ways. In the case of HfO$_2$, we will show this changes the identification of which structure is up and which is down. We will then discuss the implications of this for the interpretation of observations of piezoelectricity in ferroelectric HfO$_2$, and in particular offer a natural explanation for the apparent discrepancy between theoretical predictions of negative piezoelectric response and experimental observations in which most systems show a positive piezoelectric response~\cite{Liu20p197601,Dutta21p1,Schenk20p1900626,Boscke11p102903,Starschich14p202903,Chouprik19p275}.
Based on these results, we propose a combined theory-experiment approach to identify the preferred polarization switching path in double-path ferroelectrics by computing the piezoelectric response for a given variant and identifying it as up or down based on the shape of the measured butterfly piezoelectric hysteresis loop.

\begin{figure}[hpbt]
\includegraphics[width=7.5cm]{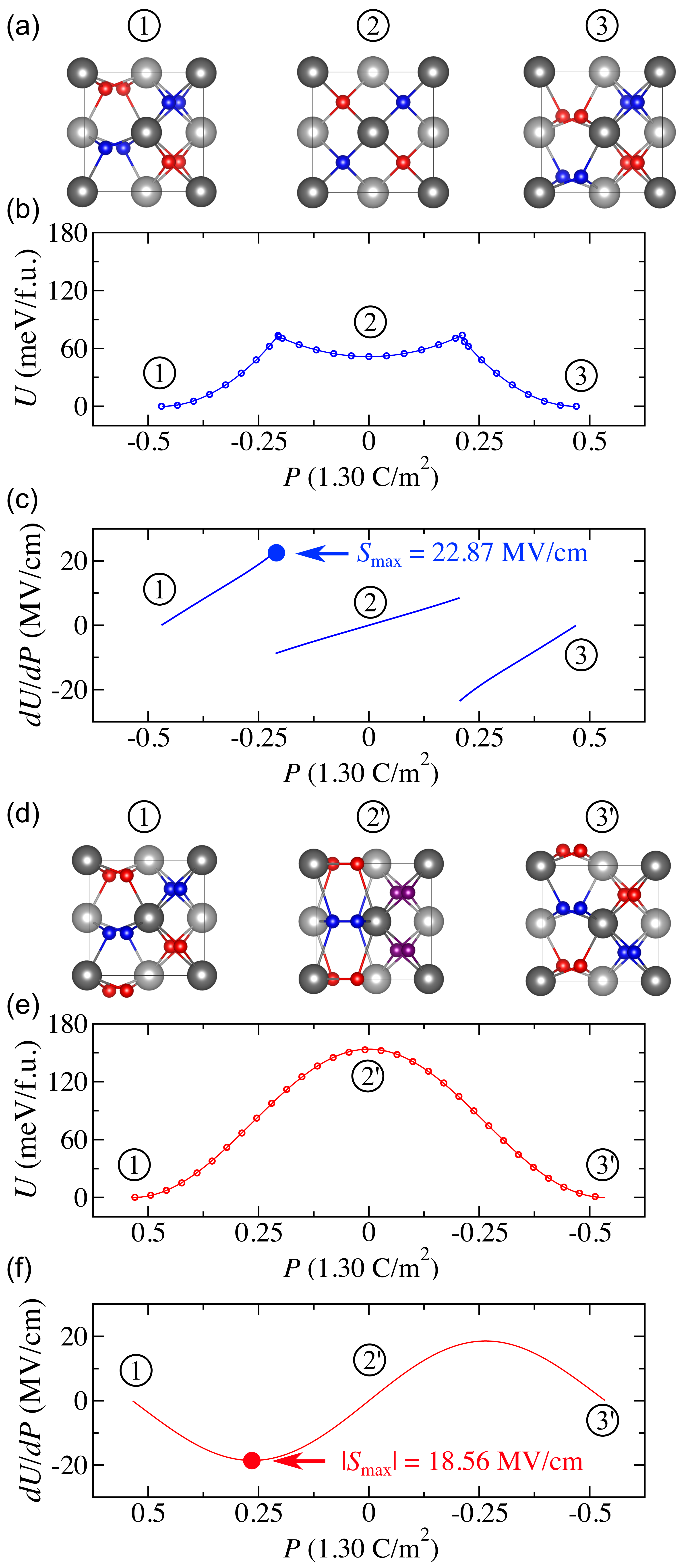}
\caption{Structure sequence ((a) and (d)), energy profiles ((b) and (e)), and slopes of the energy profiles ((c) and (f)) in the two polarization switching paths. In (a) and (c), the gray atoms are Hf and the blue and red atoms are oxygen, with red being closer and blue being farther from this viewpoint.
In the two paths, the starting structures (structure 1) are identical, with extra oxygen atoms shown in (d) to clarify the different structural evolution along path II, and the final structures (structures 3 and 3') are symmetry-related. The initial sign of $S$ is an indication that structure 1 is the up-polarized state for path I and the down-polarized state for path II. The maximum magnitudes of $S$ for paths I and II are marked in (c) and (f), respectively.}
\label{f2}
\end{figure}

The ferroelectricity in HfO$_2$-based materials is attributed to the formation of the $Pca2_1$ phase~\cite{Sang15p162905,Boscke11p102903,Qi20p257603,Xu21p826,Raeliarijaona21p1}, which is shown in Fig.~\ref{f2}.
The identification of a particular variant as up or down may at first seem as obvious as in BaTiO$_3$ or PbTiO$_3$, with polar states generated by freezing in a polar lattice mode of the high-symmetry cubic fluorite structure (Fig. 2 (a)). Since the polarization change from {1} to {3} is positive, this identifies structure {1} as down and structure {3} as up~\cite{Yang20p064012,Huan14p064111,Maeda17p1,Fan19p21743,Choe21p8,Chen20p085304,Ding20p556,Lee20p1343}. 
However, we note that in contrast to conventional ferroelectrics, additional modes are needed to generate the polar structure of HfO$_2$ and in fact there are no unstable polar modes in the the cubic fluorite structure. 
Furthermore, there is a distinct polarization switching path that has recently been discussed~\cite{Clima14p092906,Clima20p1,Yang20p064012}.
In this path~\rom{2}, shown in (d), the oxygen atoms move up through the Hf atomic planes, with the polarization change then being positive. This changes the identification of structure {1} to the up state, with structure {3'} being the down state, noting that structure {3'} is symmetry-related to structure {3}~\cite{Qi21p1}.

In this work,
we carry out density functional theory (DFT) calculations with the \textsc{Quantum--espresso} package~\cite{Giannozzi09p395502etalp} to investigate the competition between the two paths,. 
Norm-conserving pseudopotentials based on the local density approximation are generated by the \textsc{Opium} package~\cite{Opium}. 
The atomic force convergence threshold is set as 1$\times$10$^{-4}$ Hartree per Bohr. 
Calculations are done in the conventional 12-atom cell with
a 4$\times$4$\times4$ Monkhorst--Pack $k$-point mesh to sample the Brillouin zone~\cite{Monkhorst76p5188}. 
To generate the energy profile as a function of polarization, starting from structure 1, we move all oxygen atoms together along the $z$ direction,
then optimize the structure keeping the $z$-direction displacement between the Hf atom and the average position of its surrounding 8 oxygen atoms fixed~\cite{Reyes14p140103,Qi20p214108}, and then compute the polarization for the optimized structure.
The polarization of each structure is calculated with the 
Berry flux diagonalization method reported in Ref.~\cite{Bonini20p045141}.

The calculated energy profiles are shown in Fig.~\ref{f2}, with the energy of the $Pca2_1$ phase taken as zero.
In path~\rom{1}, the zero polarization structure is the centrosymmetric tetragonal $P$4${\rm{_2}}/nmc$ phase, obtained by freezing the unstable $X_2^{-}$ phonon into the cubic fluorite structure~\cite{Huan14p064111,Yang20p064012}. This tetragonal $P$4${\rm{_2}}/nmc$ phase has no unstable phonons~\cite{Reyes14p140103,Qi20p257603}.
On this path, as the polarization increases from structure {1} and reaches a critical value ($P=0.206$ C/m$^2$), 
the structure becomes unstable and collapses in a first-order phase transition to an $Aba2$ structure, which is related to the $P$4${\rm{_2}}/nmc$ phase by a polar distortion~\cite{Reyes14p140103,Qi20p214108}.
In path~\rom{2}, the structures and energies change smoothly. 
The zero polarization structure on this path is the centrosymmetric orthorhombic $Pbcm$ structure, in which the oxygen atoms lie in the Hf atomic planes~\cite{Clima14p092906,Clima20p1,Yang20p064012,Chen20p085304}.
This orthorhombic $Pbcm$ structure is unstable and has a higher (153 meV {\textit{vs.}} 51 meV per formula unit) energy than the centrosymmetric tetragonal $P$4${\rm{_2}}/nmc$ structure in path~\rom{1}.

Since path~\rom{1} has a lower energy barrier and a more symmetric zero-polarization structure,
it has been generally accepted as the physically relevant path~\cite{Yang20p064012,Huan14p064111,Maeda17p1,Fan19p21743,Choe21p8,Chen20p085304,Ding20p556,Lee20p1343}.
However, the physical quantity determining which path is favored in electric-field switching is not the height of the energy barrier, but the minimum electric field needed to drive the system through the path.
This is given by the maximum slope of the energy profile $S=\partial{U}/\partial{P}$, shown in Fig. 2 (b) and (e) for the two paths being considered here.
In path~\rom{1} (Fig. 2 (b)), $S$ increases and reaches its maximum value right before the first-order phase transition, after which $S$ jumps down and changes its sign.
The structures and energies along path~\rom{2} evolve smoothly with changing polarization, as shown in Fig. 2 (d).
The maximum slope of path~\rom{2} is in fact a little smaller than that in path~\rom{1} (18.8 MV/cm \textit{vs.} 22.9 MV/cm), making path~\rom{2} slightly more favorable. 
However, the difference is small so that the two paths should be considered competitive, with the preferred path depending on the details of the system (doping, sample preparation, temperature, stress, and bonding condition with electrodes)~\cite{Clima14p092906,Yang20p064012}.

\begin{figure}[hpbt]
\includegraphics[width=7.5cm]{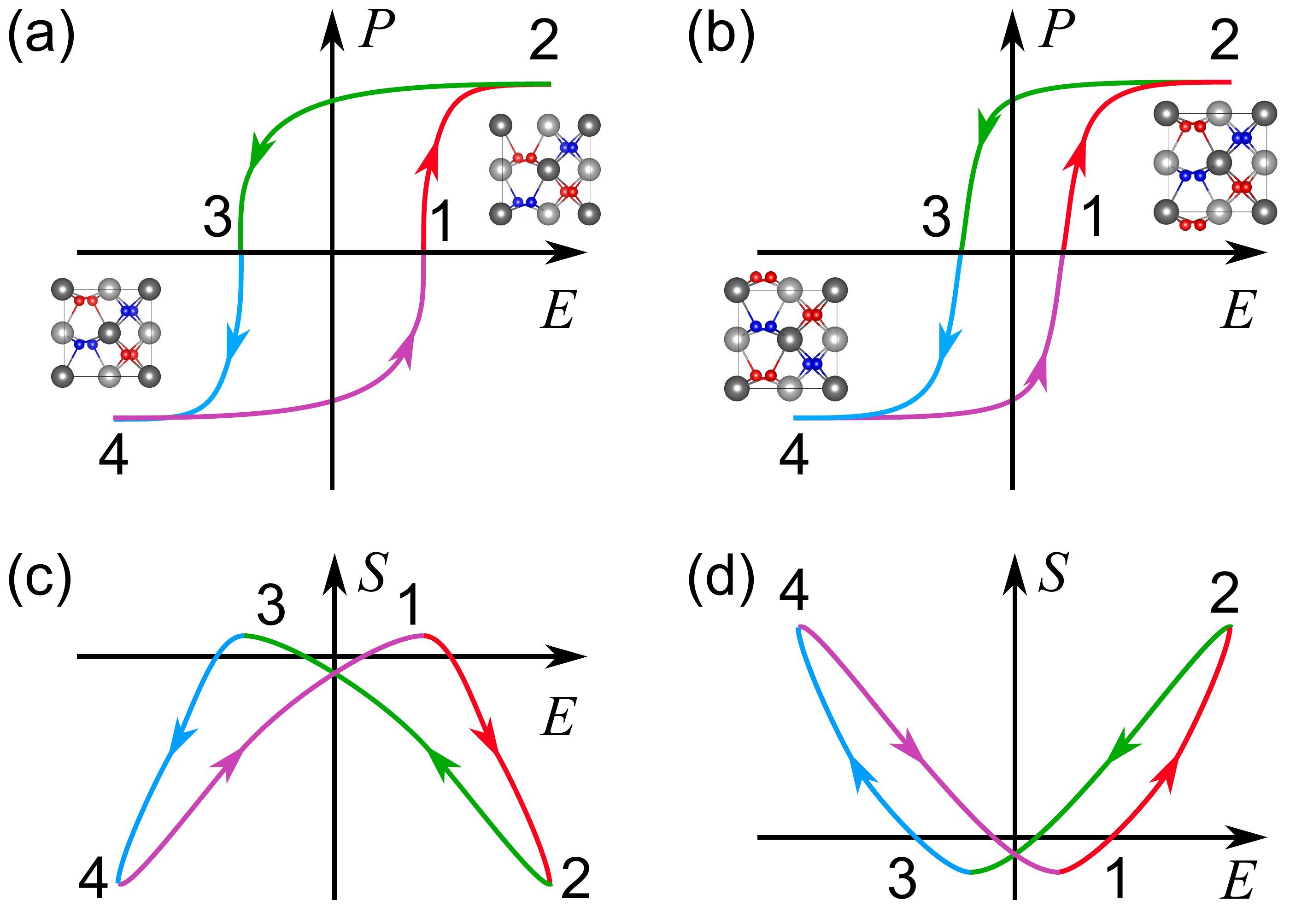}
\caption{Schematic illustrations of the electrical and piezoelectric hysteresis loops of HfO$_2$-based ferroelectrics switching according to path I, with a negative piezoelectric response ((a) and (c)) and path II, with a positive piezoelectric response ((b) and (d)). We show path I with a larger critical field than path II to schematically indicate its slightly larger maximum slope.}
\label{f3}
\end{figure}

The identification of a particular variant as ``up-polarized'' determines the sign of the piezoelectric coefficient, since the piezoelectric response of the up state is, by convention, reported as the piezoelectric response of the material.
The sign of the piezoelectric coefficient thus can reverse if a different variant is identified as ``up-polarized''~\cite{Dutta21p1}.
We first consider the case that structure 3 has been identified as the ``up-polarized'' state, as it has been in most discussions in the literature ~\cite{Liu20p197601,Yang20p064012,Huan14p064111,Maeda17p1,Fan19p21743,Choe21p8,Chen20p085304,Ding20p556,Lee20p1343}. 
First-principles calculations have shown when a uniaxial stress is applied stretching the system in the $z$ direction, the oxygen atoms move up, decreasing the polarization~\cite{Dutta21p1}. This decrease in polarization corresponds to the negative piezoelectric coefficient previously reported in first-principles studies~\cite{Liu20p197601}. 
On the other hand, if the system switches according to path II and structure 1 is identified as the ``up-polarized'' structure, a symmetry transformation of the same calculation shows that the oxygen atoms move down, increasing the polarization, and the piezoelectric coefficient is positive.
This connection between the sign of the piezoelectric response and the switching path is manifest in the experimental determination of the piezoelectric response from the piezoelectric hysteresis loop, as shown in Fig.~\ref{f3}.
If the hysteresis loop is butterfly-shaped, with positive slopes at electric field above the critical field, the piezoelectric response is positive~\cite{You19p3780};
on the other hand, if the hysteresis loop is inverted-butterfly-shaped, the piezoelectric response is negative~\cite{You19p3780}.

Positive piezoelectric responses in HfO$_2$-based ferroelectrics have been observed in many experimental works.
The samples include but are not limited to Si-doped HfO$_2$, Y-doped HfO$_2$, and La-doped HfO$_2$~\cite{Schenk20p1900626,Boscke11p102903,Starschich14p202903}.
This discrepancy with the previous first-principles prediction has attracted considerable interest and
Dutta {\textit{et al.}} showed that the computed sign of piezoelectric response of pure HfO$_2$ can become positive under a 6~\% compressive strain.  
Here, we provide an alternative explanation which is natural and straightforward;
the observed sign of the piezoelectric response can be understood by identifying the up-polarized state in the experiment as having the opposite polarization to that assumed in the previous first-principles studies.

Further, the fact that some samples show negative piezoelectric response~\cite{Dutta21p1,Chouprik19p275} can also be readily understood.
HfO$_2$ has two competing polarization switching paths with approximately equal preference. 
Therefore, it is highly plausible that polarization switching in different experiments could occur along different paths, leading to both positive and negative reported piezoelectric responses.
As an example, we consider the effect of substitution of Hf by Zr.
In Fig.~\ref{f5}, we plot the slopes of energy profiles of Hf$_{0.25}$Zr$_{0.75}$O$_2$ (75~\% Zr doped HfO$_2$) given by our DFT calculations. 
Comparing these results to those for pure HfO$_2$,
we observe that the maximum slope of path~\rom{1}
becomes smaller relative to path~\rom{2}, leading to a theoretical prediction that Zr doping will favor path~\rom{1} and a negative piezoelectric response.
In fact, Chouprik \textit{et al.} has reported an ``anomalous'' piezoresponse force microscopy (PFM) switching, which is an indication of negative piezoelectric response, in Hf$_{0.5}$Zr$_{0.5}$O$_2$ thin films~\cite{Chouprik19p275}, consistent with the theoretical prediction.

\begin{figure}[hpbt]
\includegraphics[width=7.5cm]{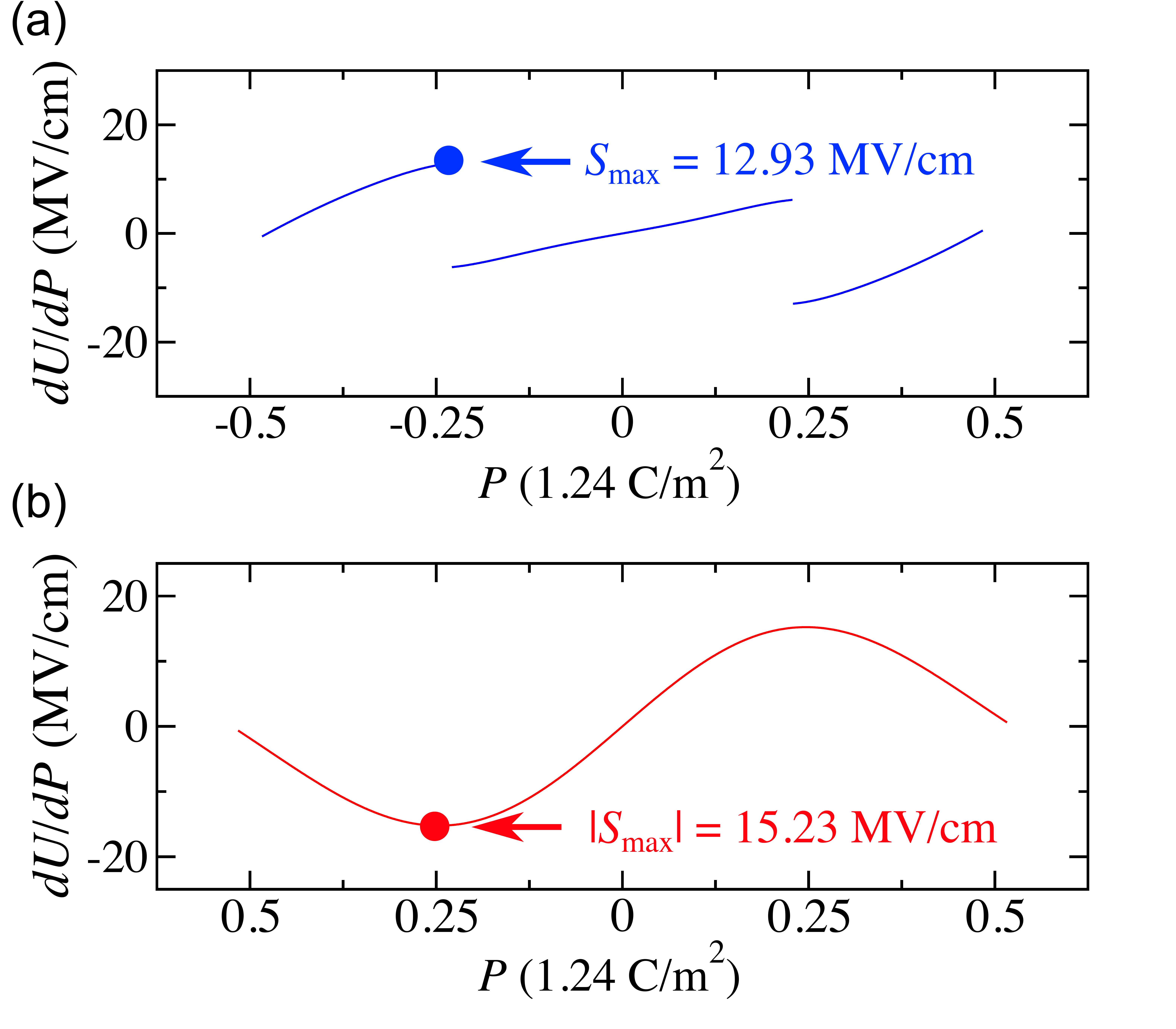}
\caption{The slopes of the energy profiles for (a) path~\rom{1} and (b) path~\rom{2} of Hf$_{0.25}$Zr$_{0.75}$O$_2$ (75~\% Zr doped HfO$_2$).}
\label{f4}
\end{figure}

We should note that our analysis so far is based on uniform switching in pure HfO$_2$, which is typically different from that of real ferroelectrics.
Polarization usually switches through a domain wall nucleation and growth process, rather than uniformly~\cite{Merz54p690,Stadler63p3255,Tybell02p097601,Ahn04p488,Li04p1174,Gruverman05p082902,Shin07p881,Liu16p360}.
Determining the details for a specific system, which involves domain wall motion~\cite{Grimley18p1701258,Buragohain18p222901,Mulaosmanovic17p3792,Fengler17p1600505,Stolichnov18p30514}, temperature~\cite{Zhou15p240,Park18p1700489}, vacancies~\cite{Schenk14p19744,Baumgarten21p032903}, surfaces/interfaces~\cite{Park17p9973}, doping~\cite{Pevsic16p4601,Hoffmann15p072006,Schroeder14p08LE02,Park17p4677,Mueller12p2412,Muller11p114113,Olsen12p082905} and so on,
is much more complex and beyond the scope of this work. However, our main interest is in the total displacement of each ion from the initial state to the final state, which is expected to be uniform even for inhomogeneous switching.
However, if the uniform switching paths are competitive, it is highly possible that different inhomogenous switching paths are also competitive and one can be more favorable under specific experimental conditions, while the other is more favorable under other conditions.

\begin{figure}[hpbt]
\includegraphics[width=8.5cm]{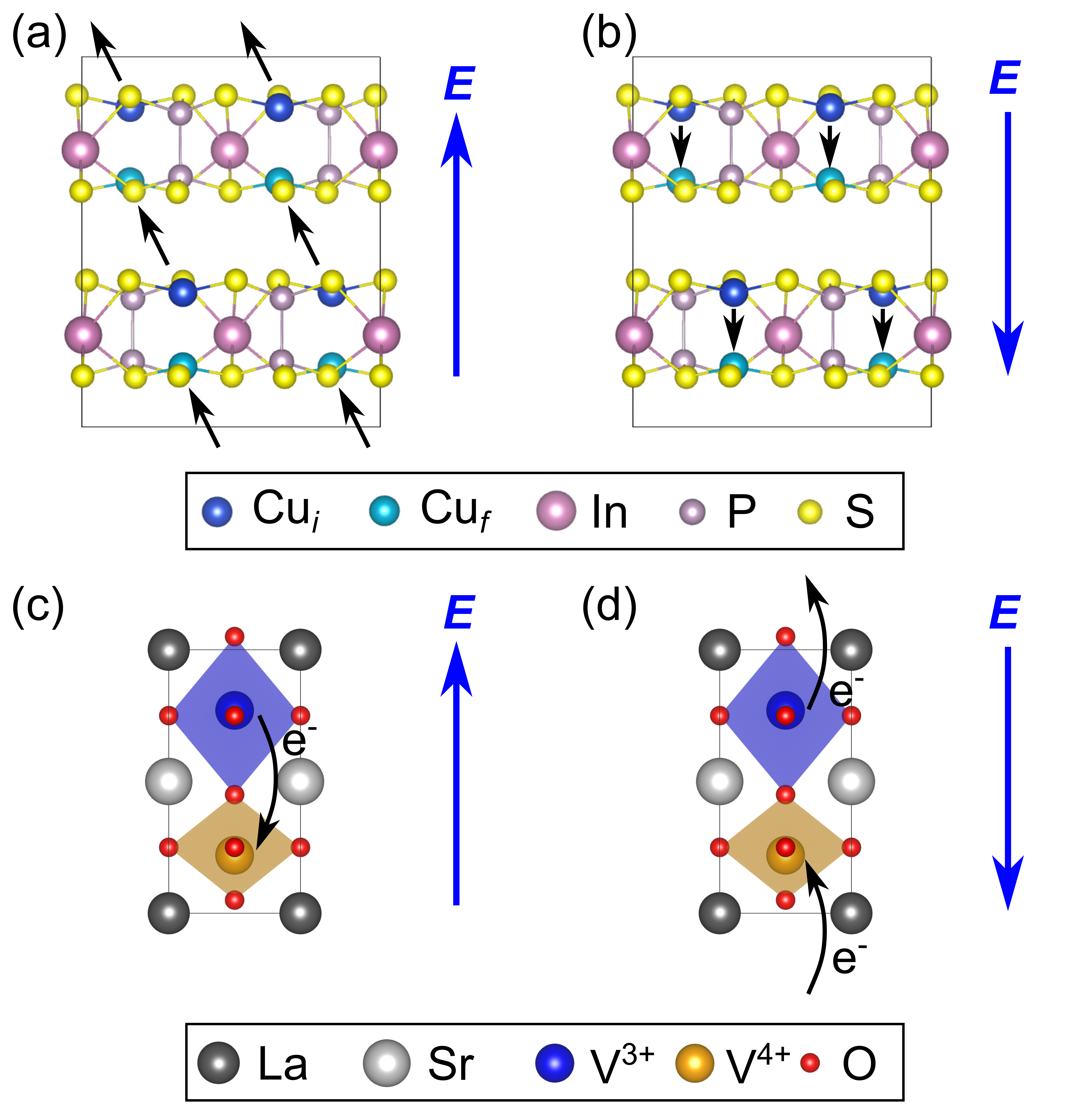}
\caption{Structures of CuInP$_2$S$_6$ and the LaVO$_3$-SrVO$_3$ superlattice, 
which are other examples of ``double-path'' ferroelectric.
In subfigures (a) and (b), the Cu atoms in the initial and final structures are represented with blue and cyan spheres.}
\label{f5}
\end{figure}

We have focused on HfO$_2$ is an example of a double-path ferroelectrics, but there are
other candidate for materials belonging to this family of ferroelectrics.
In Fig.~\ref{f5} (a) and (b), we show the structure of CuInP$_2$S$_6$ (CIPS), 
which is a two-dimensional layered ferroelectric~\cite{Neumayer19p024401,You19p3780,Brehm20p43,Neumayer20p064063}.
Each molecular layer is composed of sulfur octahedra, 
and each sulfur octahedron is filled with a Cu atom, In atom, or a P-P dimer~\cite{Liu16p12357,Maisonneuve95p157,Maisonneuve97p10860}. 
The ferroelectricity results from the displacements of the Cu atoms.
%, which are displaced from the mid-plane of the molecular layer or the gap between layer.
It has been shown that the polarization could switch either by the Cu atoms displacing across the inter-layer gap (as shown in Fig.~\ref{f4} (a))~\cite{Liu16p12357,Neumayer20p064063} or inside the molecular layer (as shown in Fig.~\ref{f4} (b))~\cite{Neumayer20p064063}, leading to the assignment of the variant shown as either up or down, respectively.
The LaVO$_3$-SrVO$_3$ superlattice, which is a theoretically proposed charge-ordering induced ferroelectric~\cite{Park17p087602,Qi21CO}, is another example.
In this system, the V atoms disproportionate into V$^{3+}$ and V$^{4+}$.
The superlattice has a polar metastable layered charge ordering (LCO) structure which can be stabilized over the competing structures by strain or an applied electric field~\cite{Park17p087602,Qi21CO}.
The polarization switching is associated with an electron transfer from the V$^{3+}$ ions to V$^{4+}$ ions. 
It is possible that the electron could transfer either though the SrO layer (as shown in Fig.~\ref{f4} (c)) or through the LaO layer (as shown in Fig.~\ref{f4} (d)), leading to the assignment of the variant shown as either up or down, respectively.

In experiments, the most straightforward way to identify up and down variants is determining the atomic arrangements in the field-switched up and down variants by atomic-scale imaging methods. However, this might not be possible, particularly in thin films.  
Here, we propose a combined theory-experiment approach, based on the assumption that first-principles calculations are generally quite successful in predicting the piezoelectric coefficients of a crystal~\cite{Fu00p281,Wu05p037601,Saghi98p4321,Bellaiche00p7877,Grinberg04p144118,Hamann05p035117,Cohen08p471,Wu05p037601} and that in experiments, the sign of piezoelectric response can be straightforwardly and unambiguously measured by the shape of the piezoelectric hysteresis loop~\cite{You19p3780}.
If the butterfly shape is normal (positive piezoelectric coefficient) then the variant with positive piezoelectric coefficient as determined by first-principles calculations is the up-polarized state. If the butterfly shape is reversed (negative piezoelectric coefficient), then the variant with negative piezoelectric coefficient is the up polarized state.

For most known materials, piezoelectric responses are positive, indicating that the polarization in a positively polarized state tends to increase under a tensile strain.
Exceptions with negative piezoelectric responses have attracted intensive research interest. 
Known negative piezoelectric materials include a variety of $ABC$ ferroelectrics~\cite{Liu17p207601}, several ~\rom{3}-\rom{5} zinc blende compounds~\cite{Bellaiche00p7877}, and most low-dimensional piezoelectrics~\cite{Kim19p104115,Qi21p217601}, and the search is still going on.
``Double path'' ferroelectrics not only have negative piezoelectric response under suitable conditions, but have the novel feature that tuning can change the sign of their piezoelectric response from negative to positive or vice versa.
%These results also indicate that double-path ferroelectrics exhibit piezoelectricity with tunable sign.
%Here, we would like to point out that the electromechanical properties of double-path ferroelectrics are even more novel;
%The piezoelectric response can be negative and switched to positive under specific conditions. 

In this work, we have proposed a specific family of ferroelectrics (referred to as ``double-path'' ferroelectric) with two competing polarization switching paths, leading to different assignments of up- and down- polarized states. 
We emphasize that the physical quantity determining the preference of each path should be the maximum slope of the energy profile, rather than the height of the energy barrier.
Examples of double-path ferroelectrics include but are not limited to HfO$_2$, CuInP$_2$S$_6$, and theoretically proposed LaVO$_3$-SrVO$_3$ superlattice. 
%In double-path ferroelectrics, the preference of one path over the other can be subtle (such as in HfO$_2$) and easily overturned by external factors, such as domain wall motion, temperature, vacancies, surfaces/interfaces, and so on. 
Double-path ferroelectrics exhibit novel electromechanical properties, 
since their piezoelectric responses can be tuned between positive and negative under different conditions.

\section*{acknowledgement}
Y.Q. and K.M.R. was supported by ONR N00014-21-1-2107. 
S.E.R.-L. acknowledges support from ANID FONDECYT Regular grant number 1220986.
First-principles calculations were performed using the computational resources provided by the Rutgers University Parallel Computing (RUPC) clusters and the High-Performance Computing Modernization Office of the Department of Defense.

\bibliography{cite}

\end{document}